\documentclass[epsfig,useAMS,usegraphicx,usenatbib]{mn2e}
\usepackage{amssymb}
\usepackage{rotate}
\usepackage[pdftex]{rotating}
\usepackage{lscape}
\usepackage{txfonts}
\newcommand{\sax}{{\it BeppoSAX}}

\newcommand{\xmm}{{\it XMM-Newton}}
\newcommand{\xte}{{\it Rossi X-ray Timing Explorer}}
\newcommand{\integral}{{\it INTEGRAL}}
\newcommand{\chandra}{{\it Chandra}}
\newcommand{\source}{4U~0513--40}
\newcommand{\degree}{\ensuremath{^\circ}}

\title[]{The 17 min orbital period  in the Ultra Compact X-ray Binary 4U~0513-40
}
\author[]{M. Fiocchi$^{1}$\thanks{E-mail:
mariateresa.fiocchi@iasf-roma.inaf.it; 
},
A. Bazzano$^{1}$,  L. Natalucci $^{1}$, R. Landi$^{2}$, P. Ubertini$^{1}$
\\
$^{1}$INAF/IASF-Roma, Via Fosso del Cavaliere 100, I-00133, Roma, Italy\\
$^{2}$INAF/IASF-Bologna, Via P. Gobetti 101, I-40129 Bologna, Italy\\
}

\begin{document}


\date{Accepted . Received ; in original form }

\pagerange{\pageref{firstpage}--\pageref{lastpage}} \pubyear{}

\maketitle

\label{firstpage}

\begin{abstract}
 The ultracompact low-mass X-ray binary \source\/ in the globular cluster NGC1851 exhibits large
amplitude X-ray flux variations with spectral changes from low/hard to high/soft states
which have not been reported previously in other ultracompact X-ray binaries.
Using \sax\/, \chandra\/ and \xmm\/ archival data together with recent \integral\/ observations, we
reveal a clear  sinusoidal periodic signal with a period of $\sim$ 17 minutes when the source is in a typical high/soft state with a dominant soft thermal component.
The periodicity disappears when the source is in a low/hard state and the thermal soft component is not required any more to model the data.
These properties indicate the orbital nature of the detected signal and imply an high inclination angle 
of the binary system ($>80\degree$).

\end{abstract}

\begin{keywords}
gamma-rays: observations X-rays: observations
\end{keywords}

\section{Introduction}
Ultracompact X-ray binaries (UCXBs) are systems with orbital periods ($P_{\rm orb}$)
shorter than $\approx$1~hr in which a neutron star or black
hole accrete matter from a companion low mass star.
 Their short periods
rule out ordinary hydrogen-rich companion stars, since these stars are
too big and do not fit in the Roche lobe (Nelson
et al. 1986).
UCXBs are rare objects and their identification is very difficult
because of the difficulty to measure $P_{\rm orb}$
in LMXBs.
The most recent compilation of ultracompact X-ray binaries lists 27 candidates (in 't Zand et al.,  2007).
Eight out of 52 LMXBs with measured orbital period are
in the ultracompact regime
(in 't Zand et al. 2007, Nelemans \& Jonker 2006). 
  The remainder were
classified as ultracompact X-ray binaries on the base of some
combination of tentative orbital period measurements, deep optical
spectra lacking hydrogen emission lines, high ratios of X-ray to
optical flux, or persistent emission at low fractions of the Eddington
rate (in 't Zand et al. 2007 and references within).
The IBIS results of the long monitoring of the UCXBs  showed that these
sources spend most of the time in the canonical low/hard state,
with X-ray luminosities $\lesssim~7\times10^{36}~erg~s^{-1}$,  plasma temperature
$kTe \gtrsim~20$ keV and $\tau~\lesssim~4-5$ (Fiocchi et al. 2008).\\
4U~0513-40 is an X-ray binary in the Galactic
globular cluster NGC~1851 with a 17-minute orbital modulation first observed with the {\em Hubble Space Telescope} (Zurek et al. 2009).
It is a persistent source showing
evidence for variability of a factor of $\sim10$ in X-ray luminosity
on timescales of $\sim$ weeks, and a factor of more than 20 overall (Maccarone et al. 2010).
Such variability is unusual for ultracompact X-ray binaries and hence
deserve some attention.
\begin{figure*}
\includegraphics[angle=-90, width=0.65\linewidth]{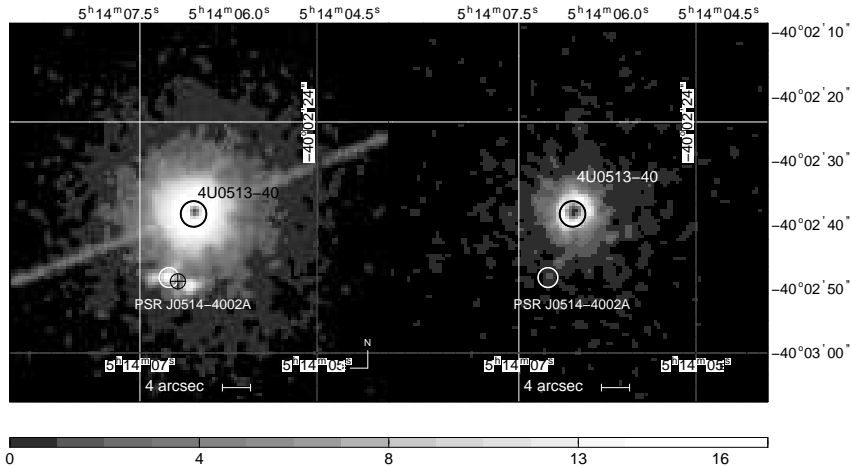}
\caption{Left: ACIS 0.5$-$8 keV image of the region surrounding \source\/.
A Gaussian smoothing was applied to the counts distribution with a width of 2 pixels.
The black circle represent the detection of the X-ray source \source\/, the cross indicate the pulsar PSR J0514-4002A position and the white circle is the \chandra\/ detection.
Right: ACIS 4$-$8 keV image of the region surrounding \source\/.
A Gaussian smoothing was applied to the counts distribution with a width of 2 pixels.
The black circle represent the detection of the X-ray source \source\/.
}
\label{two}
\end{figure*}
\section{Observations and Data Analysis}
Table \ref{jou} gives a summary of source observations performed with instruments on board \sax\/, \chandra\/, \xmm\/ and \integral\/ satellites.

The LECS, MECS and PDS/\sax\/ event files and spectra were generated with the Supervised Standard Science Analysis (Fiore, Guainazzi \& Grandi 1999). Both LECS and MECS spectra were accumulated in circular regions of $8\arcmin$ radius. Publicly available matrices were used for these instruments. 
The PDS spectra were extracted using the XAS version 2.1 package
 (Chiappetti \& Dal Fiume 1997). The background sampling was performed by making use of the 
 default rocking law of the two PDS collimators that sample ON/+OFF, ON/-OFF fields for
 each collimator with a dwell time of 96 s (Frontera et al. 1997). When one collimator is 
 pointing ON-source, the other collimator is pointing toward one of the two OFF positions. 
 We used the standard procedure to obtain PDS spectra (Dal Fiume et al. 1997).
The \chandra\/ data were processed 
with the CIAO (\textit{Chandra} Interactive Analysis of Observations) software, version 4.1.2, i.e. the 
same version of CALDB (Calibration Data Base), provided by the \textit{Chandra}
X-ray Center and following the science threads listed on the CIAO
website\footnote{Available at http://cxc.harvard.edu/ciao/}.
The CIAO routine {\ttfamily wavdetect} was used to search for X-ray sources on the ACIS chips.
 The CIAO routine {\ttfamily dmextract} was used to
produce energy spectra and {\ttfamily mkacisrmf} and {\ttfamily mkarf} for the response and ancillary files respectively.
We extracted source photons from a circular region centered on the source
with an extraction region of 8 arcsec. For the background, we used circular source-free regions in the same CCD of the studied source.\\
\xmm\/ data have been processed starting from the observation 
files with {\ttfamily SAS 7.0.0}. X-ray events
corresponding to patterns 0-4 were selected from EPIC-pn camera.
We used the most updated calibration 
files available at the time of the reduction for each source data.
Source light curves and spectra were extracted from circular regions of
10$\arcsec$ centered on the source, while background products were obtained from
off-set regions close to the source.
The ancillary and detector response matrices were generated using
the XMM-\textit{Newton} SAS {\ttfamily arfgen} and {\ttfamily rmfgen} tasks.\\
The analyzed \integral\/ (Winkler et al.\ 2003) data consist of all public observations in which \source\/ was within the field of view of the high-energy detectors. Broad-band spectra, $\sim$5--80 keV, are obtained using data from the high-energy instruments, JEM-X (Lund et al.\ 2003) and IBIS (Ubertini et al.\ 2003). The IBIS and JEM-X data have been processed using the Off-line Scientific Analysis (OSA v.\ 9.0) software released by the \integral\/ Science Data Center (ISDC, Courvoisier et al.\ 2003). Light curves and spectra are extracted for each individual science windows. These runs were performed with AVES cluster, designed to optimize performances and disk storage
for the INTEGRAL data analysis (Federici et al. 2010).
\section{Image analysis}
The excellent angular resolution provided by \chandra\/ allows us to resolve
the \source\/ X-ray emission from the binary Pulsar PRS J0514-4002.
Figure 1 shows the 0.5-8.0 keV (left panel) and 4-8 keV ACIS image (right panel).
The CIAO routine {\ttfamily wavdetect} was used to search for X-ray sources.
This routine found two sources in the 0.5-8.0 keV:  the first located at RA=05 14 06.48 and DEC=-40 02 38.8  with a positional uncertainty of 0.64 arcsec  (1-$\sigma$ statistical errors) and the second at RA=05 14 06.79  and DEC=-40 02 48.5 with positional uncertainty of 1.3 arcsec  (1-$\sigma$ statistical errors). Only UCXB \source\/ was detected
in the 4-8 keV energy range (see Figure 1).
Image analysis of the \chandra\/ observation shows that only the ultra compact binary system is emitting at high energy and this suggest
only \source\/ as the only possible counterpart of the very
high energy (\sax\/ and \integral\/) object.
In fact the Pulsar source is too weak to be the counterparts of the IBIS and PDS source.
This is in line with the peculiar nature of this binary Pulsar, being a radio steep  and very faint pulsar (Freire et al. 2007).\\
\begin{table}
\caption{Summary of the X-ray binary 4U0513-40 observations}
\label{jou}
\begin{center}
\footnotesize
\begin{tabular}{lcc} \hline \hline
Instrument&Tstart &  Exposure    \\
&MJD&  ks   \\
\hline 
BeppoSAX/LECS&      51597.6      &   31.5                         \\
BeppoSAX/MECS&      51597.6      &  73.6                      \\
BeppoSAX/PDS&       51597.6         &               34.9             \\
XMM Newton/PN&    52730.0&                    23.5        \\ 
CHANDRA/ACIS&    54560.6       &  18.8                          \\
INTEGRAL/JEM-X& 53917.6     &     102.8                       \\
INTEGRAL/IBIS&    53380.4         &    601.5                        \\
\hline
\hline
\end{tabular}
\end{center}
\end{table}
\section{Temporal analysis}
We accumulated a 16s bin light
curve in the 0.3-5 keV energy band, using the ACIS/\chandra\/ data and calculated a power spectrum over the whole
observation following the method outlined by Israel \& Stella (1996). 
The power spectrum is shown
in Figure 2 (top panel) where  a peak at 0.00099 Hz is clearly observed. The \source\/ period was obtained with an
epoch-folding technique. The best-fit period is (1004$\pm$18) s with uncertainties at 1 $\sigma$ confidence level.
Using this period value, we folded the light curves and a sinusoidal shape of the modulation was found (see Figure 2, bottom panel) with a pulsed fraction of $\sim$ 11\% (i.e. the semi amplitude of the modulation divided by the mean source
count rate).

During the \sax\/ observation, a type-I thermonuclear burst was detected at the time 51597 MJD - 15:29:55s (see par. 5 for details).
The same procedure used for \chandra\/ data was applied to 10s bin MECS/\sax\/  light
curve in the 3-5 keV energy band, with exclusion of the burst data.
The best-fit period is (1013$\pm$14) s with uncertainties at 1 $\sigma$ confidence level.
Using this period value, we folded the light curves and a sinusoidal shape of the modulation was found (see figure 3) with a pulsed fraction of $\sim$ 4\%.

We have also searched for periodicity in the \xmm\/ and \integral\/ X-ray light curves,
but no timing modulations have been detected.
\begin{figure}
\includegraphics[angle=-90, width=0.96\linewidth]{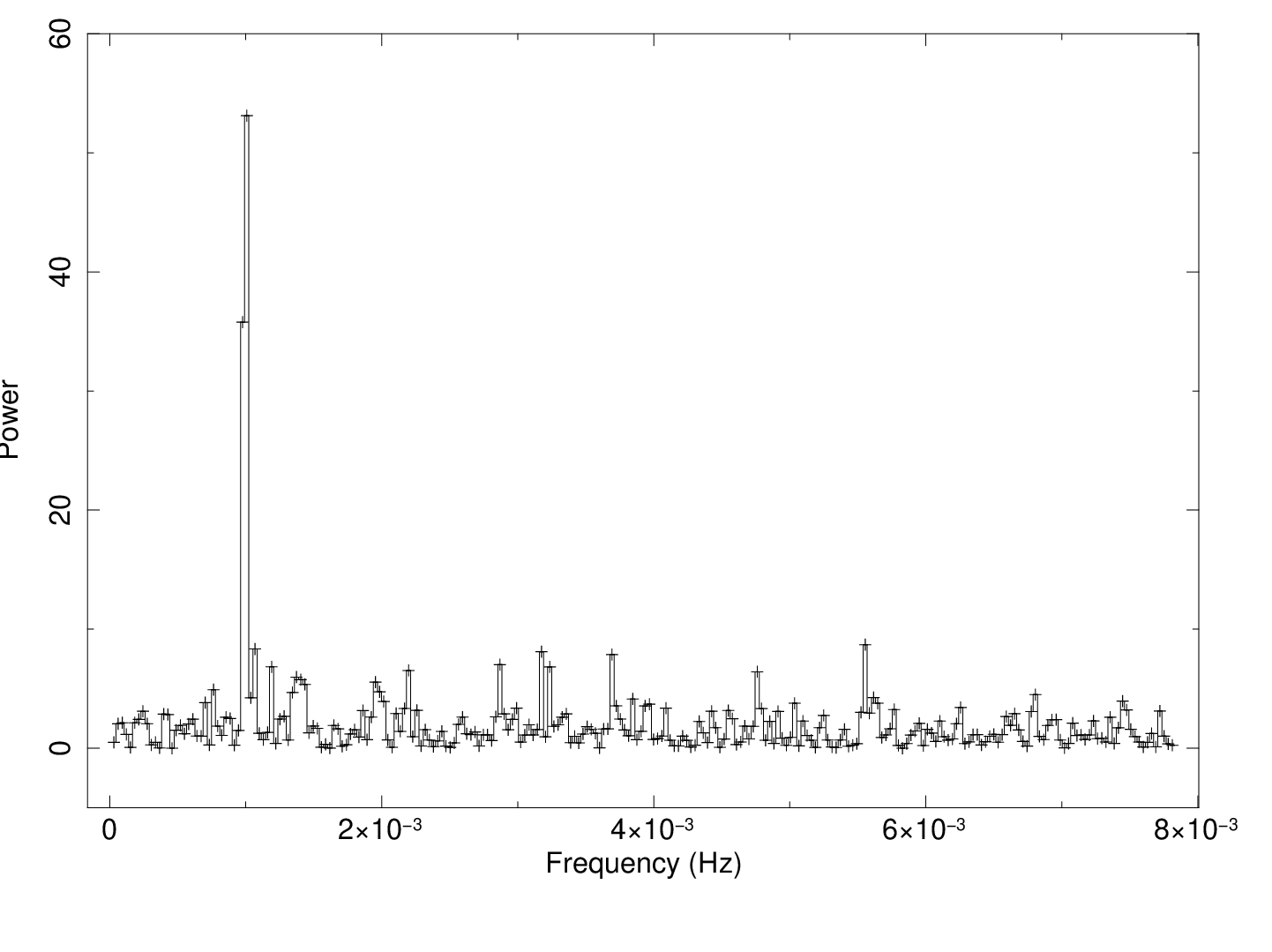}
\includegraphics[angle=-90, width=0.9\linewidth]{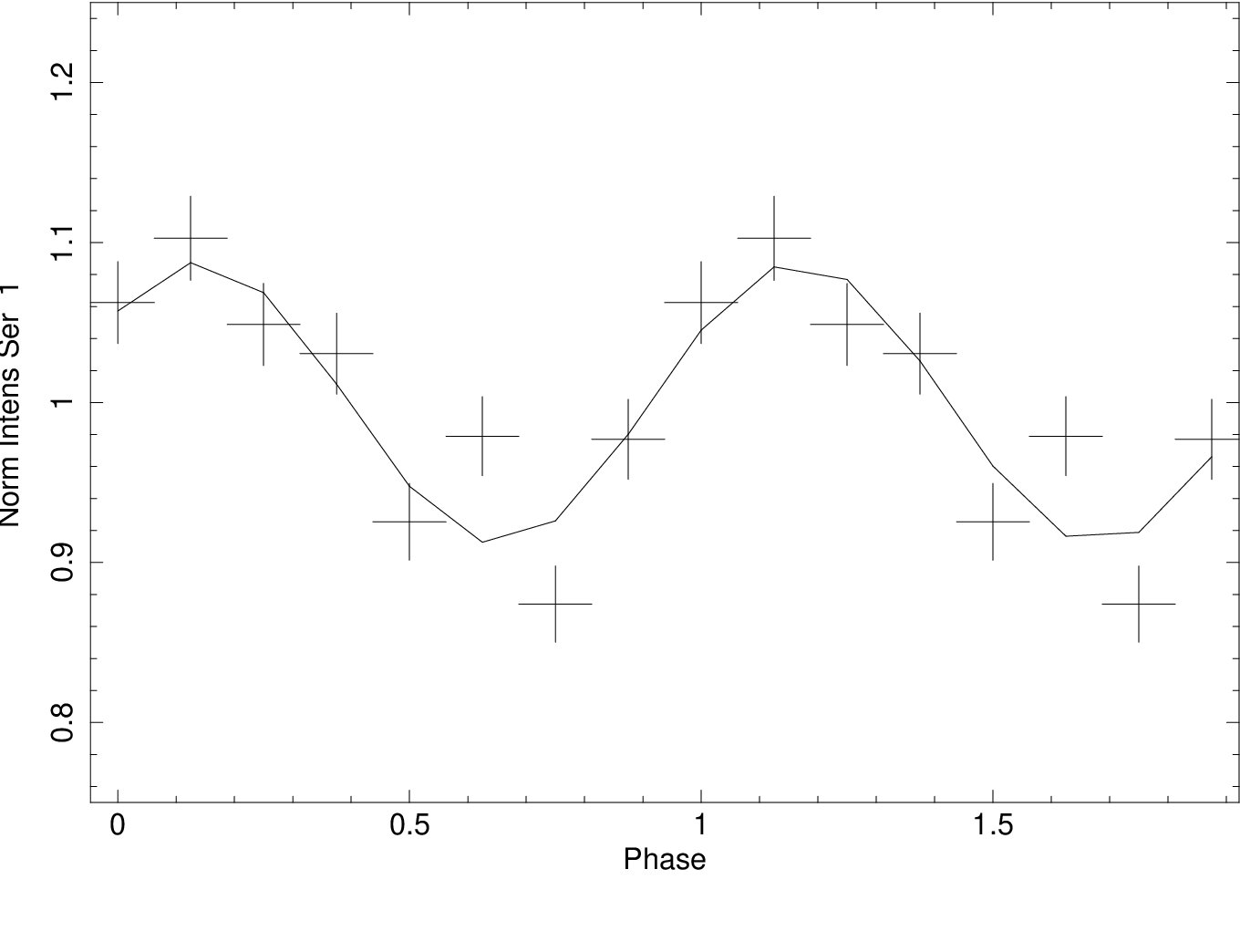}
\caption{The power spectrum and efolding phase diagram using ACIS 0.3-5 keV data
}
\label{fig3a}
\end{figure}
\section{Bursts behavior}
In the soft X-ray band, this source was observed twice with \chandra\/, once with \xmm\/ and once with \sax\/;
two thermonuclear X-ray bursts have been detected, during
the \chandra\/ and \sax\/ observations.
The first HRC/\chandra\/ burst event was described by Homer et al. (2001).
For the second not yet reported in the literature, we extract the MECS/\sax\/ light curve in the 1.5-10 keV keV energy band
and we detected an X-ray bursts starting at 51597 MJD (15:29:55 s), with a decay time of $\tau\sim$22 s, computed from exponential fits to the burst decay profile.
Fitting the spectrum (extracted at the burst peak with exposure time of $\sim$10 s) with a simple blackbody model,
a thermal temperature of 1.6$\pm$0.1keV with a extrapolated flux of 3.3$\times10^{-9}erg~cm^{-2} s^{-1}$ in 0.1-30 keV energy band has been derived.

Galloway et al. (2008) have assembled all thermonuclear type-I X-ray bursts from the accreting neutron stars observed by \xte\/, spanning on more than ten years period.
They report only 7 X-ray bursts for \source\/ and place this source
in the group of the bursters with infrequent short bursts at low accretion rate.
The sources of this sample have long orbital period (longer than 80 min) and their behavior is explained
by accreting of mixed H/He. The steady H-burning will then reduce or exhaust the accreted H prior of the bust ignition. This may not be the case
for \source\/ which has a period of 17 min.
This anomaly combinated with the detection of two bursts in the soft X-ray observation may indicate that the bursts could be very frequent and weak for this source.
\begin{figure}
\includegraphics[angle=-90, width=0.9\linewidth]{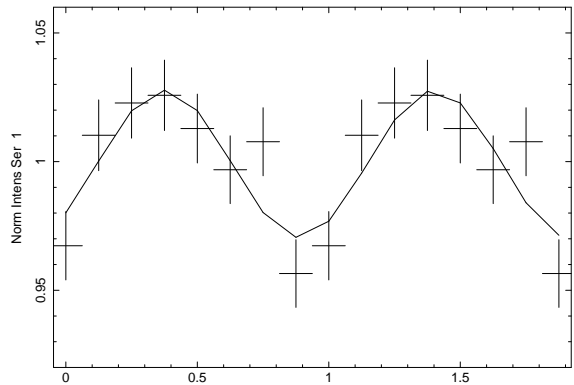}
\caption{The efolding phase using MECS 3-5 keV data
}
\label{fig3a}
\end{figure}
\section{Spectral behavior of the persistent emission}
We extracted the \sax\/ spectra before the burst event with the LECS, MECS and PDS exposure times of
1.6, 3.4 and 1.5 ks, and after the burst event with the LECS, MECS and PDS exposure times of 30.1, 67.5 and 31.2 ks, respectively.\\
IBIS and JEM-X spectra are produced summing up spectra of the source from each science windows in the period for which both instrument data are available.
The exposure times are 103 ks and 36 ks for IBIS and JEM-X, respectively.
For \chandra\/ and \xmm\/ observations, to extract the spectra we use the total exposure time, 18.8 and 23.5 ks, respectively.

Each spectrum was fitted with a model of thermal Comptonization, in XSPEC by
{\scriptsize{COMPTT}} (Titarchuk 1994; a spherical geometry was assumed), absorbed by a column density, $N_{\rm H}$.
Adding a blackbody component (modeled in XSPEC by
BBODYRAD model) resulted in a substantial fit improvement for \chandra\/ and \sax\/ (after the burst) data.
This reduces $\chi^{2}/$d.o.f.\ from 452/310 to 374/308 for the \chandra\/ data and from 173/130 to 155/128 for the \sax\/ data, with the low corresponding F-test chance probabilities of $9\times 10^{-4}$ and $4\times 10^{-13}$, respectively.

The spectral fit results are reported in Table 2 and spectra are shown in Figure 4 in different colors.
\begin{figure}
\includegraphics[angle=-90, width=1.0\linewidth]{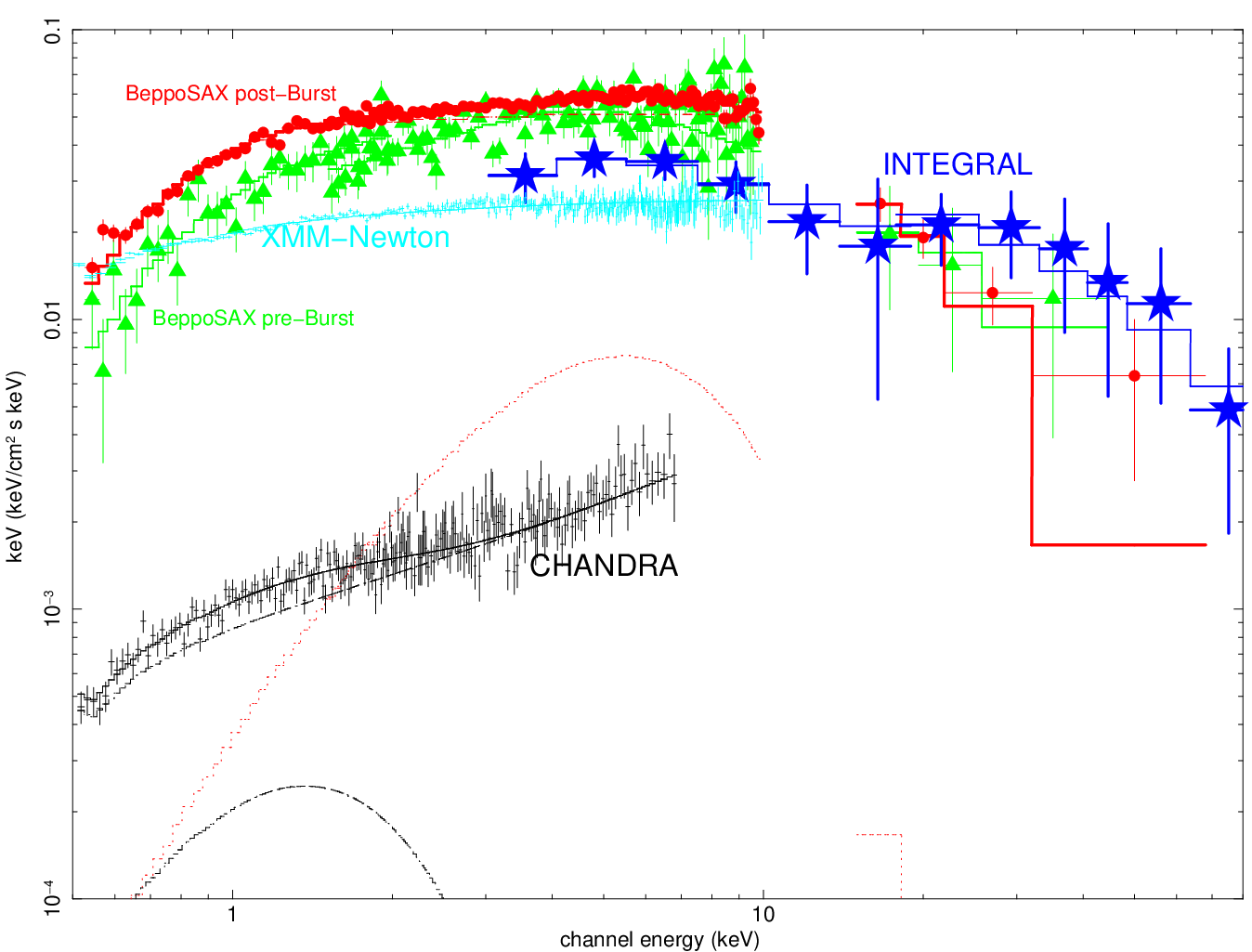}
\caption{The \sax\/, \chandra\/, \xmm\/ and \integral\/ spectra, shown together with the total model and its components.
}
\label{fig3a}
\end{figure}
\begin{table*}
\caption{Spectral analysis results. A $N_H$ fixed to the Galactic column density was included in the fit. Error are given at $90\%$ confidence level for one parameter of interest ($\Delta\chi^2=2.71$). The absorbed 1--10 keV flux are reported in units of $10^{-11}$ erg~cm$^{-2}$~s$^{-1}$.}
\label{fit}
\centering
\begin{tabular}{lcccccccc}
\hline\hline
&{kT$_{\rm BB}$}& $T_{0}$  &{kT$_{\rm e}$} &$\tau$ & $n_{\rm BB2}$ & $n_{\rm COMPTT}$ & Flux&$\chi^2$/d.o.f \\
&$keV$ &$keV$& $keV$ && &$10^{-3}$ &$10^{-11}$ erg~cm$^{-2}$~s$^{-1}$&\\
\hline
BeppoSAX spectrum before the burst event&...&$0.22\pm0.04$ &$7_{-3}^{+41}$&$3.6_{-3.3}^{+2.6}$&...&$8.6^{+15.3}_{-8.5}$ &15.2 &156/127\\
&& & & & & & &\\
BeppoSAX spectrum after the burst event&$0.49\pm0.04$& $0.18\pm0.02$&$2.6\pm0.2$&$6.8\pm0.6$&$21\pm16$ &$38\pm4$ &18.9&155/128\\
&& & & & & & &\\
\xmm\/ average spectrum&...& $0.068\pm0.002$&$15_{-10}^{+2}$&$2_{-1}^{+9}$&... &$86\pm20$ &8.9&573/387\\
&& & & & & & &\\
\chandra\/ average phase spectrum&$0.34\pm0.04$&$0.13\pm0.05$ &$5_{-2}^{+24}$&$6\pm2$&$3.9^{+2.7}_{-2.5}$ &$0.7\pm0.4$ &0.7&280/284\\
&& & & & & & &\\
\integral\/ average spectrum&...&$<1.1$ &$21_{-16}^{+32}$&$<2.4$&... &$<50$ &11.0&3/7\\
\hline
\hline
\end{tabular}\\
\end{table*}
\section[]{Discussion}
In the present paper, we have shown that the binary system 4U~0513-40 in NGC~1851
exhibits a clear periodic signal with $P \simeq 17$~min in soft X-ray.
 This signal is sinusoidal and has an amplitude from $\sim$4\% to $\sim$10\%. It is observed only in two observations
(\chandra\/ and \sax\/ after the burst) when the source is in a typical high/soft state:
the energy spectrum is well
described as the sum of a Comptonized plasma with a temperature of
$kT_e \sim$ 2-5 keV
and an optical depth of
$\tau \sim$~6 and a blackbody component with a thermal temperature of $\sim$0.3-0.5~keV.
This result is independent from the extrapolated luminosity of the system, which spans from 0.7 to 5.2 $\times~10^{36}$ erg~s$^{-1}$ in the 0.5-50 keV energy band,
for \chandra\/ and \sax\/ (after the burst), respectively.
This periodicity is not seen when the source is in a low/hard state (\xmm\/ and \integral\/ data)
and data is well reproduced by a simple Comptonized model with a plasma temperature of  $kT_e \sim$ 15-21~keV
and an optical depth of
$\tau \sim$~2, without any thermal component.
According to our present understanding,
the black-body component in the soft state could originate at both the
neutron star surface or boundary layer and the surface of an optically-thick accretion disk.
The Comptonization component in the hard state may arise from a corona above the disk and/or
between the disk and the stellar surface and accretion
probably assumes the form of a truncated outer accretion disk (Olive et al.\ 2003).
The origin of these spectral changes is not clear, the thermal temperature ($\sim$0.3-0.5 keV)
confirm that the accretion disk is ionized and should not be subject to the standard ionization instability (Done et al. 2007).

The far-ultraviolet photometry obtained with the {\em Hubble Space Telescope} has shown
the same periodicity described here (Zurek et al. 2009).
These timing properties seen in UV/optical observations 
and the eclipse observed in the X-ray band imply that the origin of this modulation is of orbital nature and the inclination angle is higher than 80 $\degree$ (Arons and King 1993).
In fact, the orbital motion  modulates the soft thermal emission coming from a small region around the neutron star but not the
 Comptonization component generated in a more extended corona above and/or around the neutron star.

\section*{Acknowledgments}
The authors acknowledge the ASI financial support
via ASI-INAF contract I/033/10/0.

\end{document}